\documentclass[10pt, a4paper]{article} 

\usepackage{xcolor}
\newcommand{\nk}[1]{\begingroup\color{orange}#1\endgroup}

\usepackage{ulem}
\usepackage{amsthm}
\usepackage{marginnote}

\usepackage[english]{babel}
\usepackage{graphicx} 

\usepackage{cite}

\usepackage[thinc]{esdiff}

\usepackage{latexsym}
\usepackage{amsfonts} 
\usepackage{amsmath} 

\usepackage{multirow}

\usepackage{enumitem}

\usepackage[utf8]{inputenc} 
\usepackage[T1,T2A]{fontenc}

\usepackage{layout}

\usepackage[left]{lineno}

\usepackage{color, soul} 
\setulcolor{yellow}

\usepackage[colorlinks=true, citecolor=blue]{hyperref}

\usepackage[nameinlink]{cleveref}  

\newtheorem{theorem}{Theorem}
\newtheorem{lemma}{Lemma}
\newtheorem{cor}{Corollary}
\newtheorem{note}{Remark}

\title{Dynamic Pricing for Real Estate}
\author{Lev Razumovskiy \and Mariya Gerasimova \and Nikolay Karenin}
\newcommand{\Addresses}{{
\bigskip
\footnotesize

Lev Razumovsky, \textsc{RAMAX Group}\par\nopagebreak
  \textit{E-mail address} : \texttt{lev.razumovskiy@ramax.com}
  
\medskip
  
  Mariya Gerasimova \textsc{}\par\nopagebreak
  \textit{E-mail address} : \texttt{marijyagerasimova@yandex.com}
  
\medskip
Nikolay Karenin, \textsc{RAMAX Group}\par\nopagebreak
  \textit{E-mail address} : \texttt{nikolay.karenin@ramax.com}

}}
\date{}

\begin{document}
\maketitle

\begin{abstract}
We study a mathematical model for the optimization of the price of real estate (RE). This model can be characterised by a limited amount of goods, fixed sales horizon and presence of intermediate sales and revenue goals. We develop it as an enhancement and upgrade of the model discussed in \cite{Besbes-12} now also taking into account variable demand, time value of money, and growth of the objective value of Real Estate with the development stage.   

\end{abstract}


\section{Introduction}

 The topic of dynamic pricing has received a considerable amount of attention in recent
 years, from different scientific communities. Broadly speaking, dynamic pricing refers to the process of determining optimal selling prices for products, goods, or services in environments where the pricing technology and infrastructure allow for frequent and cost-effective price adjustments. This is particularly relevant for companies operating online or in physical stores equipped with technologies like electronic price displays that enable regular price updates. The ability to frequently adjust prices with minimal effort and cost is a direct result of advancements in digital technologies, which have made it possible to tailor prices in response to shifting market conditions.

The study and application of dynamic pricing techniques have been extended to various business sectors, consistently demonstrating a significant increase in overall revenue. A crucial component for the success of these techniques is the availability of substantial sales data, often a natural byproduct of online commerce and, more broadly, the integration of digital technologies into retail operations.
 

Sales data collected through digital technologies offer valuable insights into customer behavior, particularly the relationship between pricing and purchase propensity.
Scientific research on methods for analyzing sales data, extracting information about how purchase propensity varies with price, and designing strategies for optimizing total revenue is both a fascinating and highly relevant theoretical field. Furthermore, mastering the implementation of efficient dynamic pricing techniques provides a substantial competitive advantage, making it a topic of significant practical importance for market players.



Dynamic pricing has generated a substantial body of literature, driven by both its theoretical significance and practical applications. Researchers from various disciplines have contributed to this field, including \textit{Operations Research}, \textit{Management Science}, \textit{Applied Mathematics}, \textit{Mathematical Modeling}, \textit{Pricing Strategies}, and \textit{Behavioral Economics}. 

Several reviews have been published that aim to compile articles on dynamic pricing from researchers across various fields \cite{Araman-11,Aviv-12,Christ-11,Chen-15,vandenBoer-15}.
In this group one can also mention the following authors: Bitran and Caldentey \cite{Bitran-03}, Elmaghraby and Keskinocak \cite{Elmaghraby-03}, Heching and Leung \cite{Heching-05}, Talluri and van Ryzin \cite{Talluri-06}, Gönsch et al. \cite{Gonsch-09}, Rao \cite{Rao-09}, Chenavaz et al. \cite{Chenavaz-11}, Deksnyte and Lydeka \cite{Deksnyte-12}, Özer and Phillips \cite{Ozer-12} and Phillips \cite{Phillips-21}.
Some reviews are more limited in scope, and report the articles on dynamic pricing models applied to specific markets: the electricity market \cite{Faruqui-10,Khan-16,Dutta-17}, ground transportation \cite{Saharan-20}, air transport \cite{Mcafee-06} to cite a few.



In this work, we consider a scenario with a limited inventory of goods and a fixed sales horizon. 
One of the first publications to discuss this scenario is that by
Kinkaid and Darling \cite{Kinkaid-62}. More recently Gallego and van Ryzin have reported on a similar scenario \cite{Gallego-13}. 
The model discussed in \cite{Gallego-13} has been used as a starting point, and then extended and modified by other authors, e.g. introducing limitations on the number of  prices or price
variations allowed \cite{Feng-95,Bitran-97,Feng-00,Feng-00b}, or introducing a non-constant demand function \cite{Feng-00c,Zhao-00}, or introducing multiple sales outlets working in parallel 
  \cite{Bitran-98} or multiple products sold in parallel \cite{Gallego-97,Kleywegt-01}. In  \cite{Besbes-12} Besbes and Maglaras additionally considered intermediate sales and revenue constraints commonly used in the real estate sector.
    
 We further improve and expand the model used in \cite{Besbes-12} by taking into account variable total demand, time value of money, and growth of the objective value of real estate with the development stage. 

 The paper is organized as follows. In Section~\ref{sec:base_model} we describe the basic model and the algorithm for finding the optimal pricing policy from \cite{Besbes-12}.
In Section~\ref{sec:variable_demand} we extend the algorithm to the case of variable demand and in Section~\ref{sec:tvm} we further improve the model by adding time value of money and growth of the objective value of Real Estate. Since a general solution can no longer be derived in a closed form, in Section~\ref{sec:objective_value} we present a solution for the case of a linear purchase propensity. 

\subsection{Main results and contributions} 



The primary contribution of this paper is the formulation of an optimization problem aimed at identifying the optimal pricing policy under conditions of variable total demand. We introduce an algorithm designed to construct this optimal pricing policy and provide a proof demonstrating its optimality.

Secondly, we extend the previous model by incorporating the time value of money and the variability of the objective RE value. We formulate optimization problems to determine pricing policies that maximize total revenue and minimize total sales (the dual problem of the former) under conditions of fixed total sales and fixed total revenue, respectively. An algorithm for deriving the optimal pricing policy is presented, and we prove the optimality of this policy in the case of linear purchase propensity function.

Finally, we conduct simulations using historical total demand to compare basic pricing policies with those that consider the time value of money and variable objective RE value. From a practical standpoint, these simulations demonstrate that any feedback policy adjusted to account for these factors significantly outperforms the unmodified feedback policy.

\section{Basic model}\label{sec:base_model}

To study the optimization of the total revenue of a RE management company, we first introduce in this section a simpler deterministic model, presented in \cite{Besbes-12}, which will be further expanded in Section \ref{sec:tvm}. 

We consider a process of sales that starts at time $t=0$ and must finish at time $t=T$. Denote by $S$ the total number of RE objects, which must be sold over this period. 
We assume that the demand process is continuous and deterministic with a rate $\Lambda(t)$.
We also assume that all the real estate objects are homogeneous in their quality. This means that the price $p(t)$ is the same for all the real estate inventory.

Let us introduce a function $v(p)$ that represents the fraction of the customer flow $\Lambda(t)$ willing to purchase an RE object for the price $p$. Function $v(p)$ can be viewed as a propensity to purchase for the price $p$. Let $p^*:=\arg \max \left(p v(p)\right)$ be the revenue-maximizing price and consider $\overline{p}=\inf \{p:v(p)=0\}\in (p^*, +\infty]$. 
We impose the following restrictions on $v(p)$:

\begin{itemize}
    \item[1.] $v(p)$ is a piecewise differentiable function.
    \item[2.] $v'(p) < 0$ for almost every $p \in (p^*, \overline{p})$.
    \item[3.] $\frac{v(p)}{v'(p)} + p$ is a non-increasing function on $(p^*, \overline{p})$.
\end{itemize}

\begin{note}
    If $v''(p)$ exists and satisfies $v''(p)\le 0$ almost everywhere on  $(p^*, \overline{p})$ then the second condition above is satisfied. Indeed,
    \begin{equation*}
        \left(\frac{v(p)}{v'(p)} + p\right)'= 2 - \frac{v(p)v''(p)}{(v'(p))^2}>0.
    \end{equation*}
\end{note}

In this model the sales rate and the revenue rate at time $t$ can be written respectively as:
\begin{eqnarray} \label{eq:sales_revenue_rates}
v(p(t)) \Lambda(t),  \\
p(t)  v(p(t)) \Lambda(t) .
\end{eqnarray}
The total sales $S(t)$ and the total revenue $R(t)$ at time $t$ can be expressed as
\begin{eqnarray}
    S(t_1)= S_{p(\cdot)}(t_1) =\int_0^{t_1}v(p(t))\Lambda(t) dt,\\ 
    R(t_1) = R_{p(\cdot)}(t_1)= \int_0^{t_1}p(t)v(p(t))\Lambda(t) dt,
\end{eqnarray}
where we sometimes will specify the pricing policy $p(t)$.
It would be convenient to use $S(t_1,t_2):=S(t_2)-S(t_1)$ and $R(t_1,t_2):=R(t_2)-R(t_1)$. 

The goal of the model is to determine the pricing policy $p_{opt}(t)$ that gives a maximum possible revenue at time $T$ and takes into account additional sales and financial milestones (constraints) one has to meet during the sales process. We consider two types of constraints: the sales targets and the revenue targets which impose a lower bound on sales and revenue acquired before some moment. Without loss of generality, we assume that both of these types of constraints are provided on a common grid $\{ 0=\tau_0< \tau_1<\ldots<\tau_k =  T\}$. The aforementioned constraints can be expressed as:
\begin{equation} \label{eq:constraints1}
S(\tau_i)=\int_0^{\tau_i}v(p(t)) \Lambda(t)dt \geq  S_i,
\end{equation}
\begin{equation}  \label{eq:constraints2}
 R (\tau_i )=\int_0^{\tau_i} p(t)v(p(t)) \Lambda(t) dt \geq R_i,
\end{equation}
where $S_i$ and  $R_i$ are the minimum amount of sales and revenue that are required at time $\tau_i$. In these notations $S_0=0$ and $R_0=0$, i.e. the constraints at $\tau_0=0$ are trivial. Also, it would be useful to consider the last sales constraint equal to the total stock $S_k=S$.   

Our problem can be now formulated as 
\begin{equation} \label{eq:main1}
S(T)=\int_0^{T}v(p(t)) \Lambda(t)dt =  S,
\end{equation}
\begin{equation}\label{eq:main2}
R(T)=\int_0^{T} p(t)v(p(t)) \Lambda(t)dt \to \max
\end{equation}
under conditions (\ref{eq:constraints1}) and (\ref{eq:constraints2}). 

We make the following feasibility assumptions throughout this paper. First, if the stock is infinite, all constraints can be satisfied by pricing policy $p(t)=p^*$ or formally:
$$\int_0^{\tau_i} v(p^*) \Lambda(t)dt \ge S_i,$$ 
$$\int_0^{\tau_i}p^* v(p^*)\Lambda(t)dt \ge R_i.$$
Second, there exists a pricing policy $p(t)$ that fulfills all constraints under the condition that the total stock is $S$. The second assumption means that there exists at least some solution to the problem and the first assumption essentially tells us that no sales constraint is too restrictive.

\subsection{Basic algorithm} \label{sec:basic}
In \cite{Besbes-12} it is shown that in the case when  $\Lambda(t) = \Lambda = \text{const}$, the optimal pricing policy $p_{\text{opt}}(t)$ is a piecewise constant function that can change only at times $\tau_i$ and only if the corresponding constraint is strictly satisfied. Let us thoroughly describe how to find this optimal pricing policy. 

The process of computing $p_{\text{opt}}(t)$ is iterative. In the first step, we start with $\tau_0=0$. At the beginning of each step, we already know $p_{opt}(t)$ until some $\tau_i$. 
We find the maximal price $p_{\tau_i}$ such that the following conditions will be satisfied if we set $p_{opt}(t)=p_{\tau_i}$ on $[\tau_i, T]$:
\begin{enumerate}
    \item all constraints will be satisfied;
    \item at least one constraint will be satisfied strictly (we will call it the most stringent constraint).
\end{enumerate}
This gives us $p_{\text{opt}}(t)=p_{\tau_i}$ on $[\tau_i, \tau_j)$, where $\tau_j$ is the time of the most stringent constraint, and we start the next step for time $\tau_j$. 

Let us describe the process of finding the most stringent constraint after time $\tau_i$ more formally. Denote by $S(\tau_i)$ and $R(\tau_i)$ already acquired sales and revenue respectively. For the sales constraint $S_k$ at time $\tau_k$ we find the price $p_{\tau_i,\tau_k}^{s,S_k-S(\tau_k)}$ such that 
\begin{equation}\label{eq:constr_price:sales}
    v\left(p_{\tau_i,\tau_k}^{s,S_k-S(\tau_i)}\right)\cdot \frac{1}{\tau_k-\tau_i}\int_{\tau_i}^{\tau_k}\Lambda(t)dt = \frac{S_k-S(\tau_i)}{\tau_k - \tau_i},
\end{equation}
i. e. the price that makes this constraint strictly satisfied. For the revenue constraint $R_k$ at time $\tau_k$ we similarly find the price $p_{\tau_i,\tau_k}^{s,S_k-S(\tau_k)}>p^*$ such that
\begin{equation}\label{eq:constr_price:revenue}
       p_{\tau_i,\tau_k}^{r,R_k-R(\tau_i)}\cdot v\left(p_{\tau_i,\tau_k}^{r,R_k-R(\tau_i)}\right)\cdot \frac{1}{\tau_k-\tau_i}\int_{\tau_i}^{\tau_k}\Lambda(t) dt = \frac{R_k-R(\tau_i)}{\tau_k - \tau_i}. 
\end{equation}
The next value of the optimal pricing policy can now be found as 
\begin{equation}
    p_{\tau_i} = \min_{k>i}\{p_{\tau_i,\tau_k}^{s,S_k-S(\tau_i)}   , p_{\tau_i,\tau_k}^{r,R_k-R(\tau_i)}\}.
\end{equation}

Note that the last factor on the left-hand side of formulas (\ref{eq:constr_price:sales}) and (\ref{eq:constr_price:revenue}) is just the average value of $\Lambda(t)$ on an interval $[\tau_i,\tau_k]$. For constant $\Lambda(t)=\Lambda$ this average value is also equal to $\Lambda$. We choose to leave formulas this way so they also can be applied to the case of non-constant $\Lambda(t)$. One can quite easily find what sales constraint is the most stringent and what revenue constraint is the most stringent. Their indices can be written as: 
$$j_S=\arg\max_{k>i}\frac{S_k-S(\tau_i)}{\tau_k - \tau_i}$$
and 
$$j_R=\arg\max_{k>i}\frac{R_k-R(\tau_i)}{\tau_k - \tau_i}.$$

The algorithm of finding the most stringent constraint is illustrated in the following figures.

\begin{figure}
    \centering
    \includegraphics[width=1\linewidth]{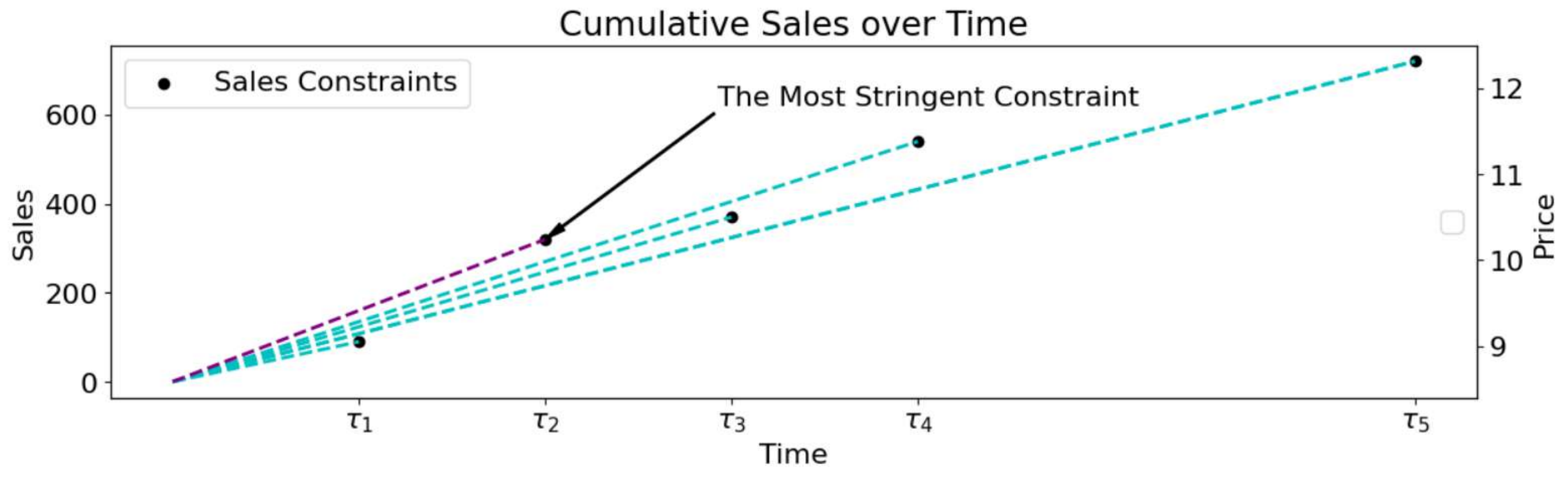}
    \caption{Finding the first most stringent constraint}
    \label{fig:stringent1}
\end{figure}

\begin{figure}
    \centering
    \includegraphics[width=1\linewidth]{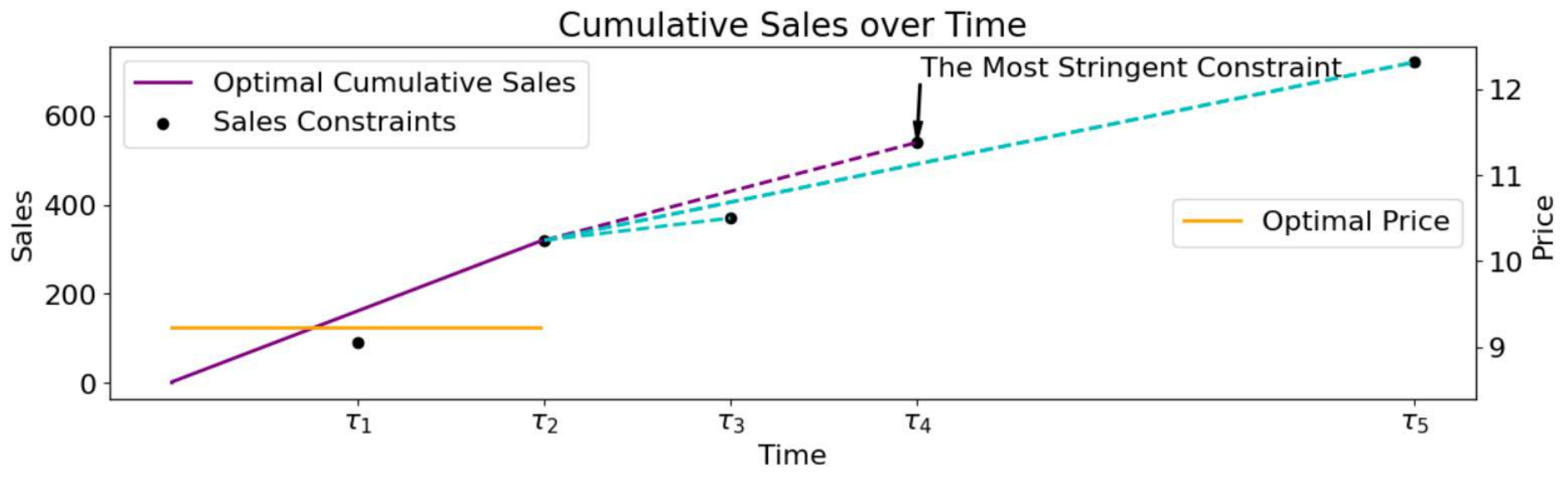}
    \caption{Finding the second most stringent constraint}
    \label{fig:stringent2}
\end{figure}

\begin{figure}
    \centering
    \includegraphics[width=1\linewidth]{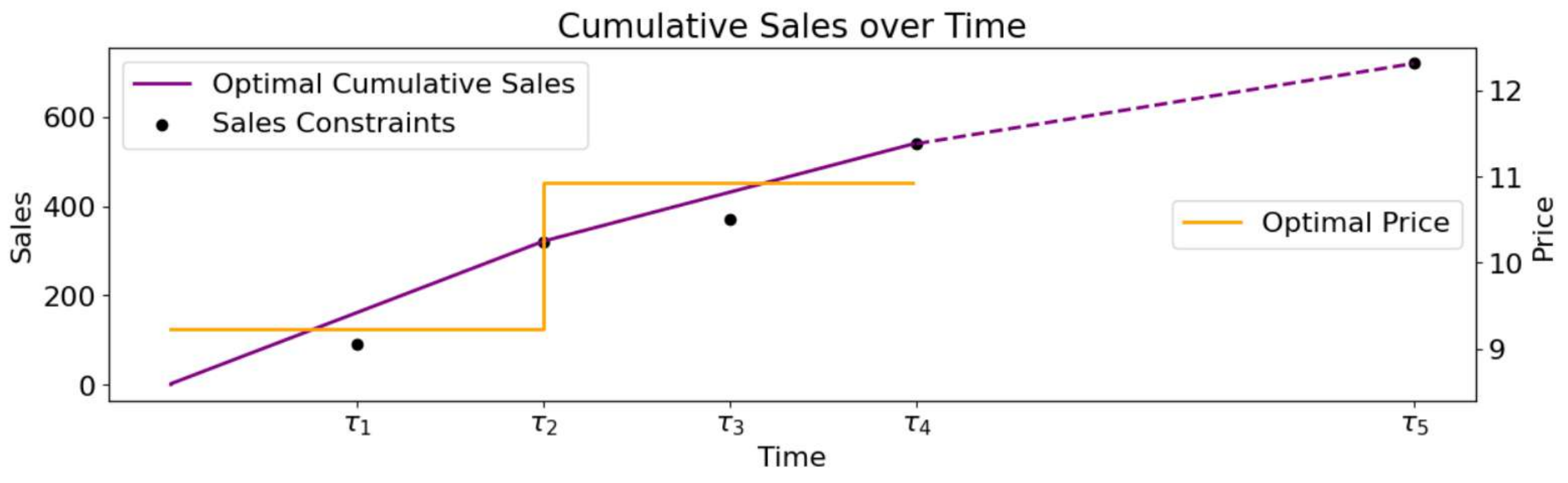}
    \caption{The last sales constraint}
    \label{fig:stringent3}
\end{figure}

\begin{figure}
    \centering
    \includegraphics[width=1\linewidth]{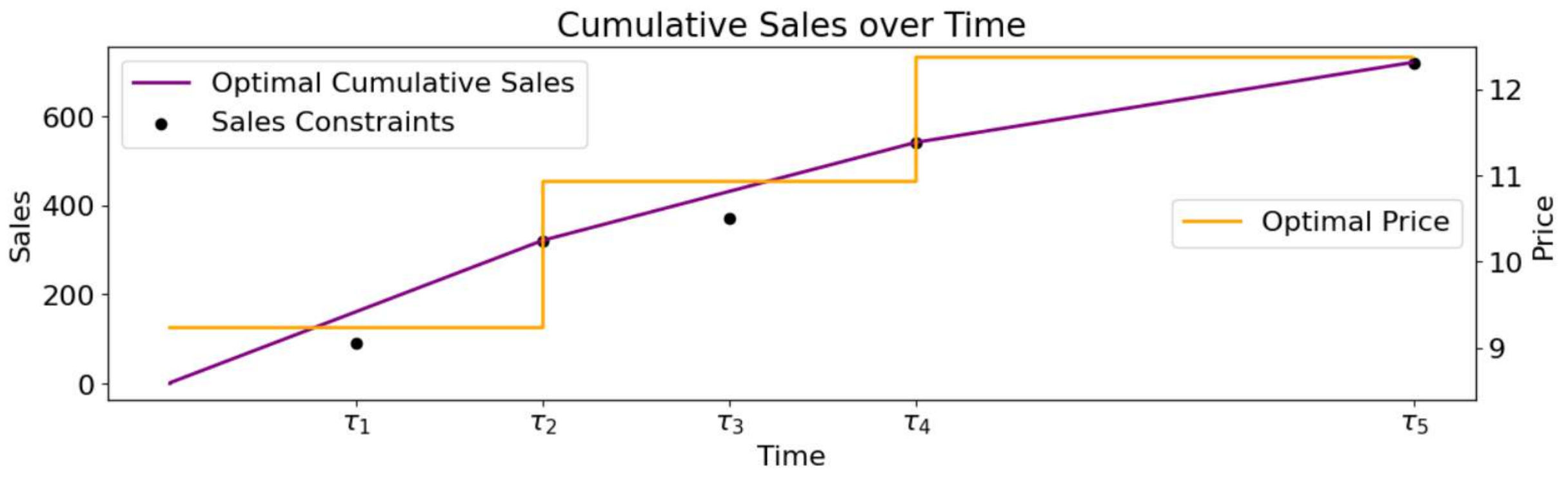}
    \caption{The optimal pricing policy and the cumulative sales}
    \label{fig:stringent4}
\end{figure}

\section{Variable total demand}\label{sec:variable_demand}


In Section \ref{sec:base_model} we discussed the basic model, that was reported in \cite{Besbes-12}, which considers the total demand rate (demand per unit time) $\Lambda(t)$ as constant. In this section, we upgrade that model and assume that the total demand is a time-varying piecewise continuous function. 

 
\begin{theorem}\label{thm1}
    Assume that $\Lambda(t)$ is piecewise continuous then the optimal pricing policy $p_{opt}(t)$ can be constructed using the algorithm from Section~\ref{sec:basic}.
\end{theorem}

Before starting the proof let us mention that this result is quite unexpected -- the optimal price does not depend on the local changes of $\Lambda(t)$ and depends only on some integral characteristics of $\Lambda(t)$ as we will show below. 

We start by proving the following lemma.
\begin{lemma}\label{lem1_1}
    Assume that there are no constraints other than the requirement to sell all the stock until time $T$. Then $p_{opt}(t)$ is constant.  
\end{lemma}
\begin{proof}
The formal statement for this problem is the following:
$$\int_0^T p(t) \Lambda(t) v(p(t)) dt \to \max,$$ 
$$\int_0^T \Lambda(t) v(p(t)) dt =S.$$
The Lagrangian functional for this problem is 
$$L(p, q)=\int_0^T p(t) \Lambda(t) v(p(t)) dt +q\cdot \left[\int_0^T \Lambda(t) v(p(t)) dt -S\right], $$
where $q$ is the Lagrange multiplier. 
Lagrange-Euler equation gives us

$$\diffp{L}{p} = p(t)\Lambda(t)v'(p(t))+\Lambda(t)v(p(t))+q\cdot \Lambda(t)v'(p(t)) = 0,$$
or equivalently 
$$\frac{v(p)}{v'(p)}+p=-q.$$
Note that the left-hand side is a monotone function of $p$, so $p(t)$ is constant. This finishes the proof.

\end{proof}
\begin{cor}\label{cor1}
$p_{opt}(t)$ is constant on the interval between any two constraints. 
\end{cor}
\begin{proof}
    The optimal price on this interval also maximizes the revenue for a given amount of sales on this interval. This happens independently of the types of constraints at the ends of the interval. It is exactly the problem considered in Lemma~\ref{lem1_1} up to a shift of time. So the optimal price is constant.
\end{proof}
\begin{proof}[Proof of Theorem~\ref{thm1}] 
Consider the pricing policy $\overline{p}(t)$ given by the description from Section~\ref{sec:basic} and assume that it differs from $p_{opt}(t)$. Both these policies are constant on the intervals between constraints. Consider $\tau_i=\inf \{t: p_{opt}(t)\ne \overline{p}(t)\}$, let $\tau_j>\tau_i$ be the time of the most stringent constraint after $\tau_i$ and let $p_0$ be the value of $\overline{p}(t)$ on the interval $[\tau_i, \tau_j)$.

First, consider the case 
\begin{equation}\label{eqn:case1}
   S_{p_{opt}(\cdot)}(\tau_i,\tau_j)<S_{p_0}(\tau_i,\tau_j).
\end{equation}
In this case, the most stringent constraint at $\tau_j$ is necessarily a revenue constraint. By Lemma~\ref{lem1_1} the optimal revenue on $[\tau_i, \tau_j)$ under the condition that sales are equal to $S_{p_{opt}(\cdot)}(\tau_i, \tau_j)$ is given by the constant pricing policy with some price $p_1$. Note that $p_1 > p_0 > p^*$ by (\ref{eqn:case1}). It follows that 
$$R_{p_0}(\tau_i, \tau_j)>R_{p_1}(\tau_i, \tau_j)\ge R_{p_{opt}(\cdot)}(\tau_i, \tau_j) \ge R_j-R_{p_{opt}(\cdot)}(\tau_i)=R_{p_0}(\tau_i, \tau_j).$$
A contradiction.

We are left with the case 
\begin{equation}\label{eqn:case2}
   S_{p_{opt}(\cdot)}(\tau_i,\tau_j) \ge S_{p_0}(\tau_i,\tau_j).
\end{equation}
By the definition of the most stringent constraint 
$$S_{p_{opt}(\cdot)}(\tau_i, T)=S-S_{p_{opt}(\cdot)}(\tau_i) < S_{p_0}(\tau_i, T),$$
hence by continuity of $S(\tau_i, t)$, there exists $t_1\in(\tau_i,T]$ such that 
$$S_{p_0}(\tau_i,t_1)=S_{p_{opt}(\cdot)}(\tau_i, t_1).$$
Consider the pricing policy 
$$
p(t)=\begin{cases}
			p_0 & \text{if } t \in [\tau_i, t_1),\\
            p_{opt}(t) & \text{otherwise}.
\end{cases}
$$
It satisfies all the constraints before $t_1$ by the definition of the most stringent constraint. Lemma~\ref{lem1_1} tells us that the pricing policy $p_0$ gives us higher revenue and the same sales on $[\tau_i, t_1)$ as $p_{opt}(t)$, so $p(t)$ also satisfies all the constraints after $t_1$ and gives higher total revenue. This is again a contradiction.
We see that our assumption was incorrect, therefore $p_{opt}(t)=\overline{p}(t)$.

\end{proof}

\section{Time value of money and variable objective RE value} \label{sec:tvm}

The Time Value of Money (TVM) is a fundamental financial principle that asserts a sum of money has greater value today than it will at a future date, due to its potential to earn returns over time and inflation. In other words, money in hand now is more valuable than the same amount received in the future because it can be invested and potentially increase in value. The Time Value of Money function $\varphi(t)$ specifies the objective value of a unit of money at time $t$.  With respect to the simple model described in Section \ref{sec:base_model}, here we discuss how to take into account the Time Value of Money. 

The TVM function modifies how we compute the revenue. Instead of computing the revenue as an integral of the revenue rate, we need to compute and maximize an objective value of the acquired revenue. For convenience, we will use the same notations that we used previously for the revenue
\begin{equation}\label{eq:tvm:revenue_at_t}
    R(t_1) = \int_0^{t_1}\varphi(t)p(t)v(p(t))\Lambda(t) dt.
\end{equation}
    The problem we consider can be formulated as follows:
\begin{eqnarray}
    R(T) \to \max, \\
    S(T) = S,\\
    R(\tau_i)\ge R_i,\\
    S(\tau_i)\ge S_i.
\end{eqnarray}
Note that here $R_i$ is the minimal objective value of revenue required at time $\tau_i$.

\subsection{Change of an objective value of the real estate objects with the development stage}\label{sec:objective_value}
Another distinctive characteristic of the real estate market is that sales often occur long before construction is completed. It is not unusual for buyers to purchase a property a year or two before it is ready for occupancy. This practice inherently involves certain risks and can be inconvenient for buyers. As a result, they tend to pay a premium for properties that are further along in the construction process compared to those at earlier stages.
This can be described mathematically using a monotonic increasing function, $\kappa(t)$, which represents the multiple by which a customer is willing to pay more for the same real estate property at time $t$ compared to what they were willing to pay for it at the start of construction under similar circumstances.

To properly take into account $\kappa(t)$ we have to consider $v(p,t)$ -- a propensity to buy a RE object that changes with time. Function $\kappa(t)$ tells us how exactly $v(p,t)$ changes:
\begin{equation}
    v(\kappa(t)p,t) = v(p,0).
\end{equation}

It will be convenient to denote a pricing policy by $\widehat{p}(t)$ and use $p(t)=\frac{\widehat{p}(t)}{\kappa(t)}$ in all the formulas. Also, we will write $v(p)$ instead of $v(p,0)$. This allows us to rewrite formulas for the sales rate and the revenue rate as 
\begin{eqnarray}
    v(\widehat{p}(t), t)\Lambda(t) = v(\kappa(t)p(t),t)\Lambda(t) = v(p(t),0)\Lambda(t) = v(p(t))\Lambda(t)\\
    \varphi(t)\widehat{p}(t)v(\widehat{p}(t),t)\Lambda(t) = \varphi(t)\kappa(t)p(t)v(p(t))\Lambda(t)=
    \zeta(t)p(t)v(p(t))\Lambda(t),
\end{eqnarray}
where $\zeta(t):=\varphi(t)\kappa(t)$ is the generalized TVM function. We get the same optimization problem as before but with $\zeta(t)$ instead of $\varphi(t)$ in formula (\ref{eq:tvm:revenue_at_t}). For the sake of clarity, we will focus only on the initial problem that takes into account $\varphi(t)$. But we will keep in mind that the same solution is applicable to this more general case.

\subsection{Maximizing revenue under the sales constraint} \label{sec:tvm:task1}

Let us first consider the problem of maximizing the revenue during the sales interval $[0;T]$ in the absence of revenue constraints and with the singular sales constraint: all RE objects must be sold by the time $T$. We have

\begin{equation} \label{eq:tvm:optimization}
	p_\text{opt}(t) = \arg \max_{p(\cdot)} \int_0^{T}\varphi(t)p(t)v(p(t))\Lambda(t) dt
\end{equation}
with the restriction on sales in the time interval $[0, T]$ being 
\begin{equation} \label{eq:tvm:total_sales}
	\int_0^{T} v(p(t))\Lambda(t) dt = S.
\end{equation}

To find the necessary maximum conditions, we will write the Lagrangian functional in a standard way:
\begin{equation}
	L(p, q) = \int_0^{T}\varphi(t)p(t)v(p(t))\Lambda(t) dt + 
 q \left[ \int_0^{T} v(p(t))\Lambda(t) dt - S \right],
\end{equation}
where $q$ is the Lagrange multiplier.
Let us write the Euler-Lagrange equation:
\begin{eqnarray}
    \frac{\partial L}{\partial p} = \varphi(t)v(p(t))\Lambda(t)+ \varphi(t)p(t)v'(p(t))\Lambda(t) +
    q v'(p(t))\Lambda(t)= 0, 
\end{eqnarray}
Since $v'(p(t))\ne 0$ we can rewrite it as  
\begin{eqnarray} 
    \frac{v(p(t))}{v'(p(t))}+p(t) =-\frac{q}{\varphi(t)} .\label{eq:tvm:result1}
\end{eqnarray} 
The solution of this equation cannot be obtained for the general case in a closed form, but one can study it for some particular types of function $v(p)$. It is important to note that for a non-constant $\varphi(t)$ the right-hand side of (\ref{eq:tvm:result1}) is also non-constant, therefore the optimal pricing policy $p_{opt}(t)$ can no longer be constant.

\subsection{The case of linear $v(p)$}
We will be mostly interested in the most commonly used $v(p)$ given by the linear formula

\begin{equation}\label{eq:linear_elasiticity}
v(p)=\begin{cases}
			1 & \text{if   } a-b\cdot p>1,\\
                0 & \text{if   } a-b\cdot p <0,\\
                a-b\cdot p & \text{otherwise}.
\end{cases} 
\end{equation}

Simple calculations show that in this model $v(p^*)=1$ if $a \geq 2$ and $v(p^*)<1$ otherwise. Since taking over the whole market while maximizing revenue is not realistic we will consider only the case $a<2$. Note that in this case $p^*=\frac{a}{2b}$ and $v(p)=a-bp$ on $[p^*,\overline{p}]$ so we can use $a-bp$ instead of $v(p)$. 

Condition (\ref{eq:tvm:result1}) gives us 
\begin{equation}
    \frac{a-b\cdot p(t)}{-b}+p(t) = -\frac{q}{\varphi(t)}
\end{equation}
or equivalently 
\begin{equation}\label{eq:tvm:price_formula}
    p(t)=\frac{1}{2}\left[\frac{a}{b}-\frac{q}{\varphi(t)} \right].
\end{equation}
We will denote a function $p(t)$ given by this formula for $q=q_0$ by $p_{q_0}(t)$. It is important to note that we consider only $q \le 0$ since $p(t)\ge p^*$. 
It will be useful to compute 
\begin{equation} \label{eq:tvm:elast_formula}
    v(p(t))=a-b\cdot p(t)=\frac{1}{2}\left[a+\frac{q\cdot b}{\varphi(t)} \right].
\end{equation}
Condition (\ref{eq:tvm:total_sales}) gives us 
\begin{equation*}
    aK(0,T) + qbI(0,T) = 2S,  
\end{equation*}
where $I(t_1,t_2) = \int_{t_1}^{t_2}\frac{\Lambda(t)dt}{\varphi(t)} $ and $K(t_1, t_2)= \int_{t_1}^{t_2}\Lambda(t)dt$.
It follows that 
\begin{eqnarray}\label{eq:tvm:q_sales}
    q = q^{s, S}_{0,T} =\frac{2S-aK(0,T)}{bI(0, T)}.
\end{eqnarray}
Here the indices indicate that $q^{s, S}_{0,T}$ corresponds to acquiring sales $S$ during the time interval $[0, T]$. 

\begin{figure}[h!]
    \centering
    \includegraphics[width=1\linewidth]{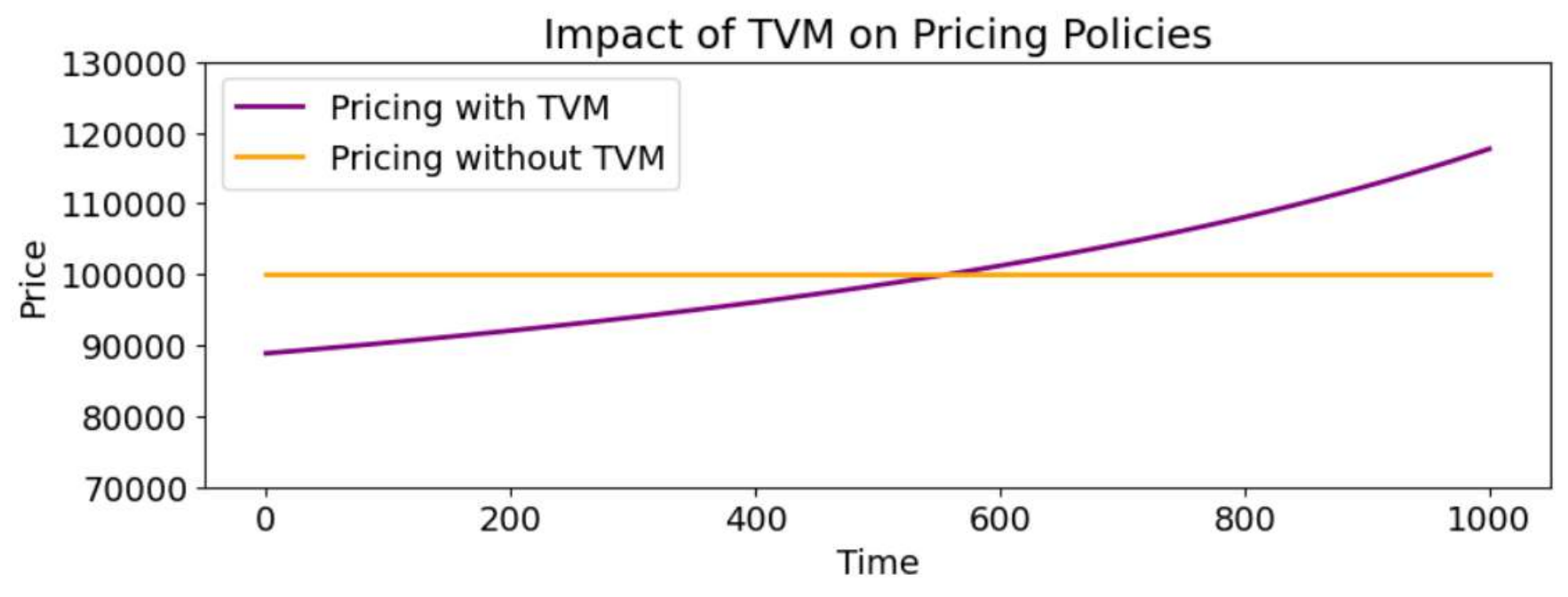}
    \caption{Comparison of two pricing policies}
    \label{fig:tvm:price}
\end{figure}

\begin{figure}[h!]
    \centering
    \includegraphics[width=1\linewidth]{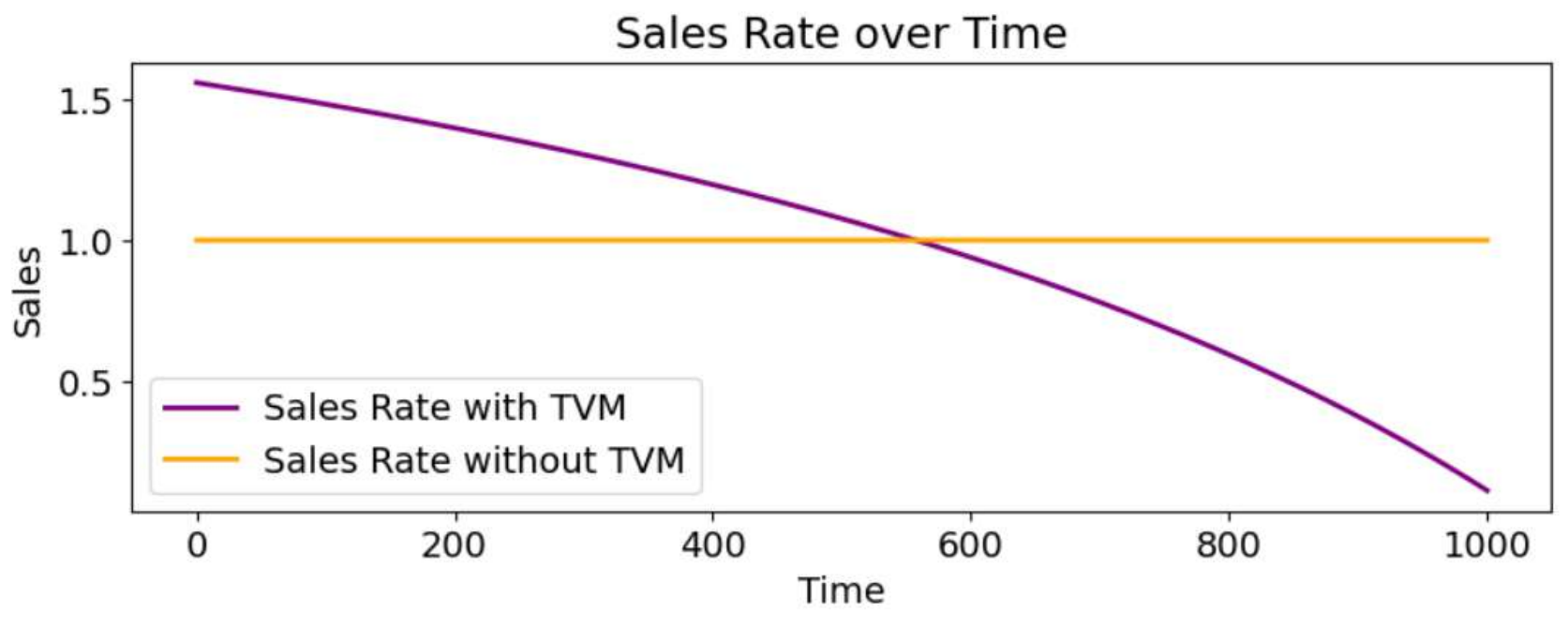}
    \caption{Comparison of two sales rate curves}
    \label{fig:tvm:sales}
\end{figure}

In figure~\ref{fig:tvm:price} we compare the purple price curve that takes into account the Time Value of Money and the yellow price curve that does not. Here for clarity we use a constant demand rate and only one sales constraint at time $T=1000$. Also we choose linearly decreasing function $\varphi(t)$. The corresponding sales curves are displayed in figure~\ref{fig:tvm:sales}. These plots completely meet our expectations -- they show that it is more profitable to sell early. 

\subsection{Minimizing sales with a given revenue value} \label{sec:task2}
Consider the dual of the problem described in Section \ref{sec:tvm:task1}, i.e. the problem of minimizing sales for a given total revenue $R$ at the final time $T$. The arguments similar to the proof of Corollary~\ref{cor1} tell us that the optimal pricing policy for this problem also satisfies (\ref{eq:tvm:result1}). The only difference is that to compute the constant $q$ one has to use 
\begin{equation*}
    \int_0^{T}\varphi(t)p(t)v(p(t))\Lambda(t) dt = R.
\end{equation*}

If $v(t)$ is linear, we can use
\begin{eqnarray*}
p(t)=\frac{1}{2}\left[\frac{a}{b}-\frac{q}{\varphi(t)} \right],\\
v(p(t))=\frac{1}{2}\left[a+\frac{q\cdot b}{\varphi(t)} \right]
\end{eqnarray*}
and get 
\begin{equation*}
    \frac{a^2}{b}J(0, T)-q^2bI(0, T)=4R,
\end{equation*}
where $J(t_1, t_2)=\int_{t_1}^{t_2}\Lambda(t)\varphi(t)dt$. Since $q<0$ we have
\begin{equation} \label{eq:tvm:q_rev}
    q = q^{r, R}_{0,T} = - \sqrt{\frac{a^2 J(0,T) - 4Rb}{b^2I(0,T)}}.
\end{equation}
Here the indices indicate that $q^{r, R}_{0,T}$ corresponds to acquiring revenue $R$ during the time interval $[0, T]$. 

\subsection{The optimal pricing policy taking into account TVM for the linear case}
We can already guess what the optimal pricing policy $p_{opt}(t)$ taking into account TVM function $\varphi(t)$ looks like. It can be constructed similarly to the optimal pricing policy constructed for the simpler model in Section~\ref{sec:basic}. 

Note a very important property of $p_q(t)$. By definition (\ref{eq:tvm:price_formula}) $p_q(t)$ is a monotone decreasing function of $q$ for any fixed $t$. This means that the minimal price corresponds to the maximal value of $q$ and vice versa.  

The process of constructing $p(t)$ is iterative. On the first step we start with $\tau_0=0$. At the beginning of each step we already know $p(t)$ until some $\tau_i$. We find the maximal $q$ from the values computed using (\ref{eq:tvm:q_sales}) and (\ref{eq:tvm:q_rev}) as 
\begin{equation}
    q_{\tau_i} = \max_{j>i}\left\{q^{s,S_j-S_{p(\cdot)}(\tau_i)}_{\tau_i, \tau_j}, q^{r,R_j-R_{p(\cdot)}(\tau_i)}_{\tau_i, \tau_j}\right\}
\end{equation}
and find the most stringent constraint -- the constraint corresponding to $q_{\tau_i}$. If the most stringent constraint is at time $\tau_j$, we define $p(t)=p_{q_{\tau_i}}(t)$ on $[\tau_i, \tau_j)$ and go to the next step. 

\begin{theorem}
    The process described above gives the optimal pricing policy $p_{opt}(t)$. 
\end{theorem}
The proof will be very similar to the proof of Theorem~\ref{thm1}.
\begin{proof}
Consider the pricing policy $\overline{p}(t)$ given by the description above and assume that it differs from $p_{opt}(t)$. 
Both these policies are given by formula (\ref{eq:tvm:price_formula}) on the intervals between the constraints. Consider $\tau_i=\inf \{t: p_{opt}(t)\ne \overline{p}(t)\}$, let $\tau_j>\tau_i$ be the time of the most stringent constraint after $\tau_i$ and let $q_0$ be the value of $q$ for $\overline{p}(t)$ on the interval $[\tau_i, \tau_j)$.

First, consider the case 
\begin{equation}\label{eqn:case1:tvm}
   S_{p_{opt}(\cdot)}(\tau_i,\tau_j)<S_{p_{q_0}(\cdot)}(\tau_i,\tau_j).
\end{equation}
In this case, the most stringent constraint at $\tau_j$ is necessarily a revenue constraint. By Section~\ref{sec:tvm:task1} optimal revenue on $[\tau_i, \tau_j)$ under the condition that sales are equal to $S_{p_{opt}(\cdot)}(\tau_i, \tau_j)$ is given by formula (\ref{eq:tvm:price_formula}) with a constant $q=q_1$. Note that $p_{q_1}(t) > p_{q_0}(t) > p^*$ by (\ref{eqn:case1:tvm}). It follows that 
$$R_{p_{q_0}(\cdot)}(\tau_i, \tau_j)>R_{p_{q_1}(\cdot)}(\tau_i, \tau_j)\ge R_{p_{opt}(\cdot)}(\tau_i, \tau_j) \ge R_j-R_{p_{opt}(\cdot)}(\tau_i)=R_{p_{q_0}(\cdot)}(\tau_i, \tau_j).$$
A contradiction.

We are left with the case 
\begin{equation}\label{eqn:case2:tvm}
   S_{p_{opt}(\cdot)}(\tau_i,\tau_j) \ge S_{p_{q_0}(\cdot)}(\tau_i,\tau_j).
\end{equation}
By the definition of the most stringent constraint 
$$S_{p_{opt}(\cdot)}(\tau_i, T)=S-S_{p_{opt}(\cdot)}(\tau_i) < S_{p_{q_0}(\cdot)}(\tau_i, T),$$
hence, by continuity of $S(\tau_i, t)$ there exists $t_1\in(\tau_i,T]$ such that 
$$S_{p_{q_0}(\cdot)}(\tau_i,t_1)=S_{p_{opt}(\cdot)}(\tau_i, t_1).$$
Consider the pricing policy 
$$
p(t)=\begin{cases}
			p_{q_0}(t) & \text{if } t \in [\tau_i, t_1),\\
            p_{opt}(t) & \text{otherwise}.
\end{cases}
$$
It satisfies all the constraints before $t_1$ by the definition of the most stringent constraint. Section~\ref{sec:tvm:task1}  tells us that the pricing policy $p_{q_0}(\cdot)$ provides a higher revenue and the same sales on $[\tau_i, t_1)$ as $p_{opt}(t)$, so $p(t)$ also satisfies all constraints after $t_1$ and provides a higher total revenue. This is again a contradiction.
We see that our assumption was incorrect, therefore $p_{opt}(t)=\overline{p}(t)$.
\end{proof}

\section{Numerical Illustrations}
To better illustrate the results of this paper, we present several simulations in this section that compare various pricing strategies.

\begin{figure}[h!]
    \centering
    \includegraphics[width=1\linewidth]{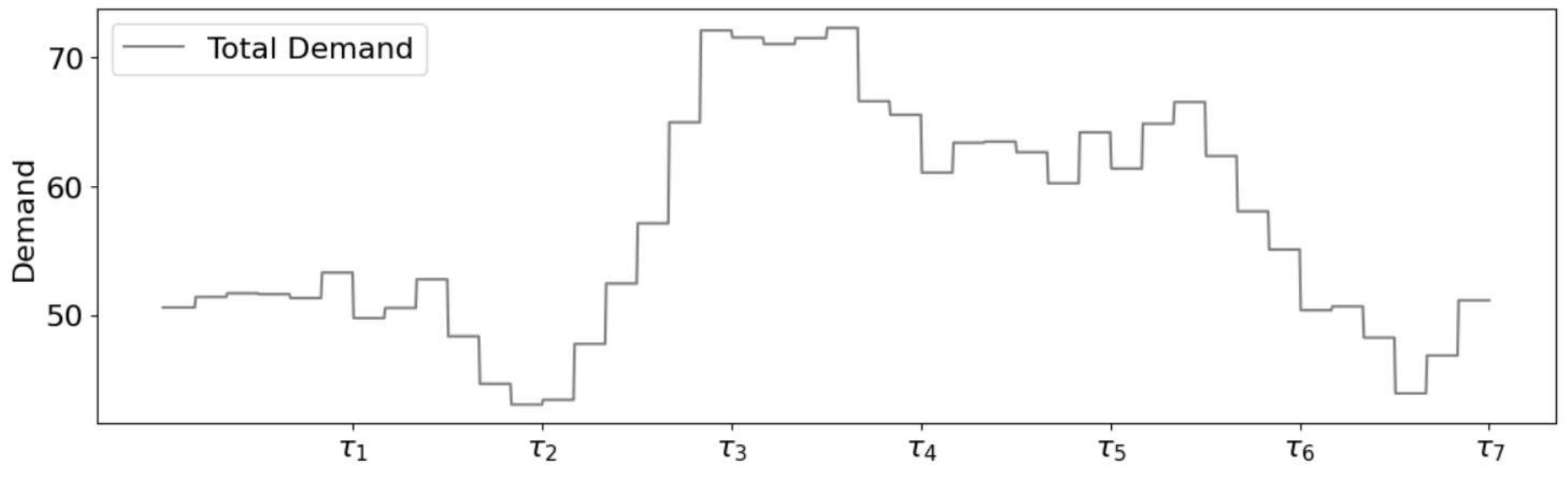}
    \caption{Total Demand}
    \label{fig:demand}
\end{figure}

In all simulations, we use historical real estate sales data from a specific city (see figure~\ref{fig:demand}) as the total demand function $\Lambda(t)$. A total of $S=1000$ objects are sold over $T=1260$ days ($3.5$ years). We assume no additional sales constraints and introduce six intermediate revenue constraints, spaced every 180 days. Each simulation compares two different strategies.

\begin{figure}[h!]
    \centering
    \includegraphics[width=1\linewidth]{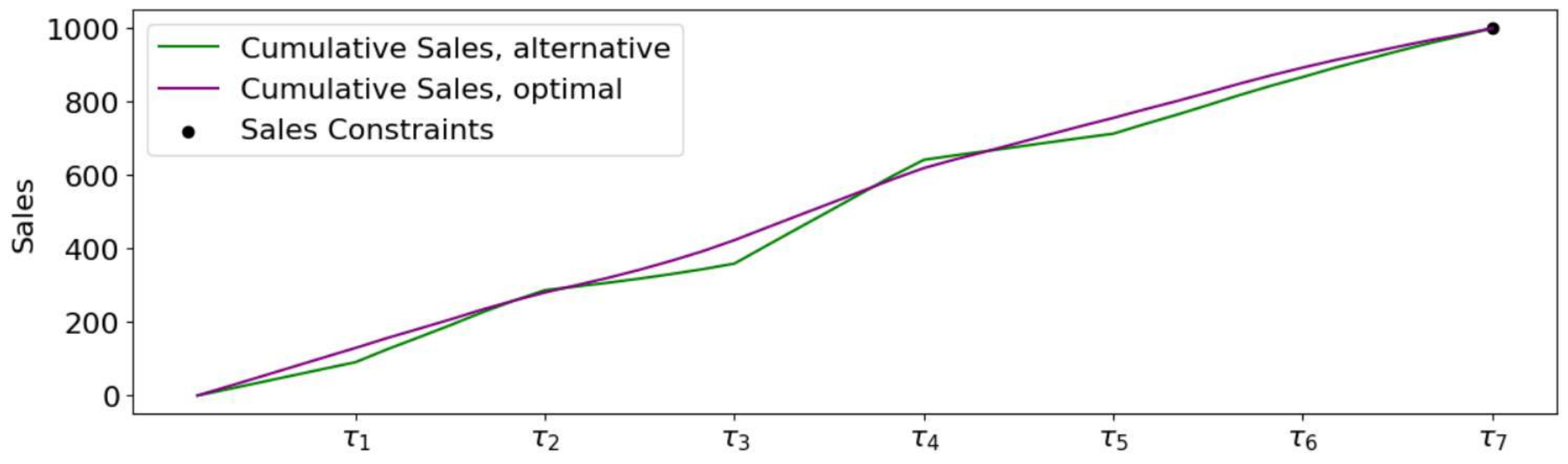}
    \caption{Simulation 1, Cumulative Sales}
    \label{fig:sim1:sales}
\end{figure}

\begin{figure}[h!]
    \centering
    \includegraphics[width=1\linewidth]{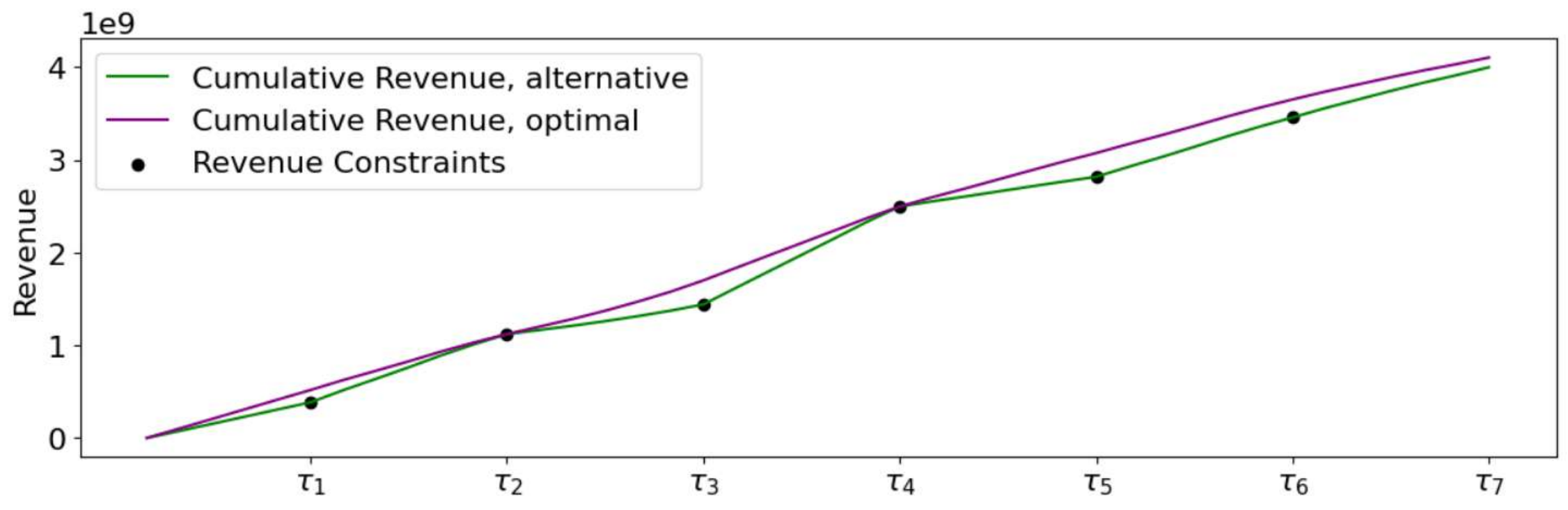}
    \caption{Simulation 1, Cumulative Revenue}
    \label{fig:sim1:revenue}
\end{figure}

\begin{figure}[h!]
    \centering
    \includegraphics[width=1\linewidth]{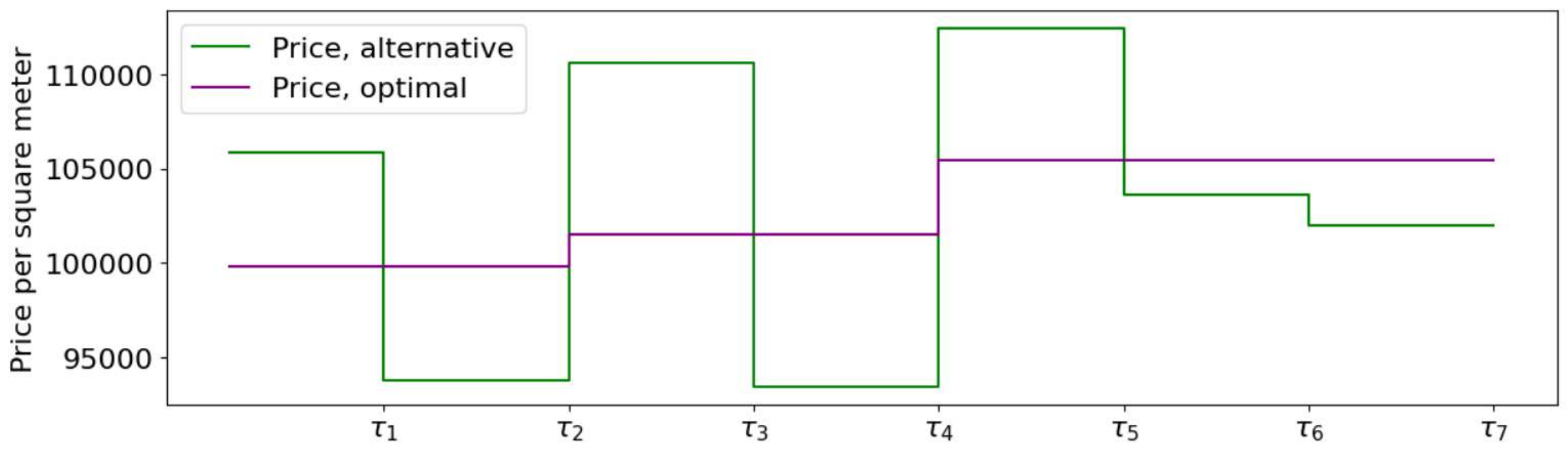}
    \caption{Simulation 1, Price}
    \label{fig:sim1:prices}
\end{figure}

In the first simulation (see figures~\ref{fig:sim1:sales},\ref{fig:sim1:revenue},\ref{fig:sim1:prices}) we compare the optimal strategy described in this paper with an alternative strategy that prioritizes satisfying the nearest constraint. In this simulation $\varphi(t)=\kappa(t)=1$. The optimal strategy yields approximately $3\%$ higher revenue. This is true in general. In general, the optimal strategy significantly outperforms any other simple strategy.


\begin{figure}[h!]
    \centering
    \includegraphics[width=1\linewidth]{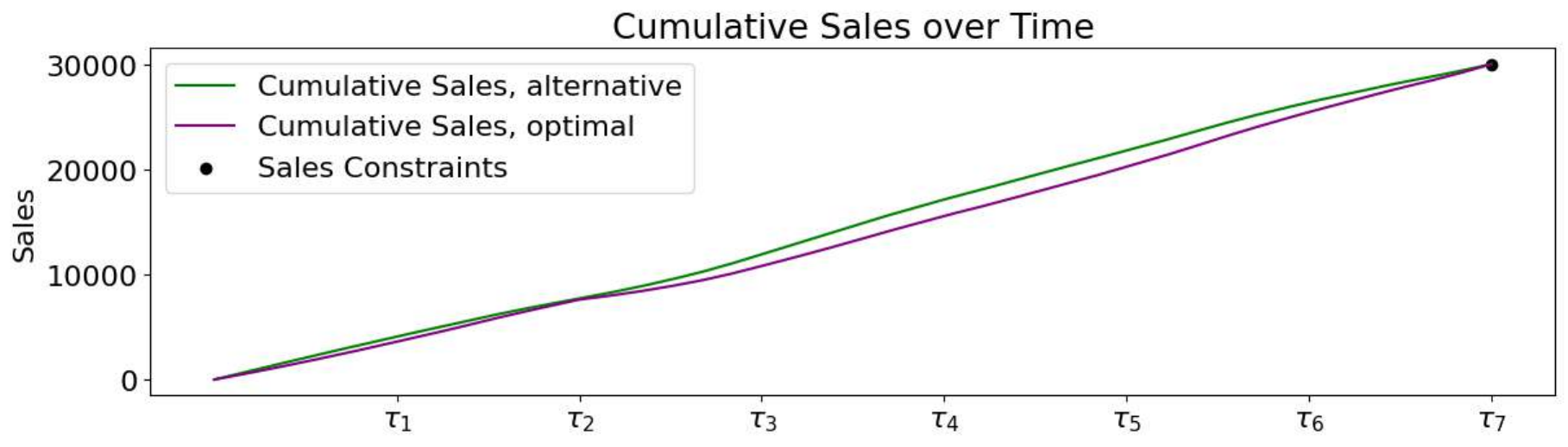}
    \caption{Simulation 2, Cumulative Sales}
    \label{fig:sim2:sales}
\end{figure}

\begin{figure}[h!]
    \centering
    \includegraphics[width=1\linewidth]{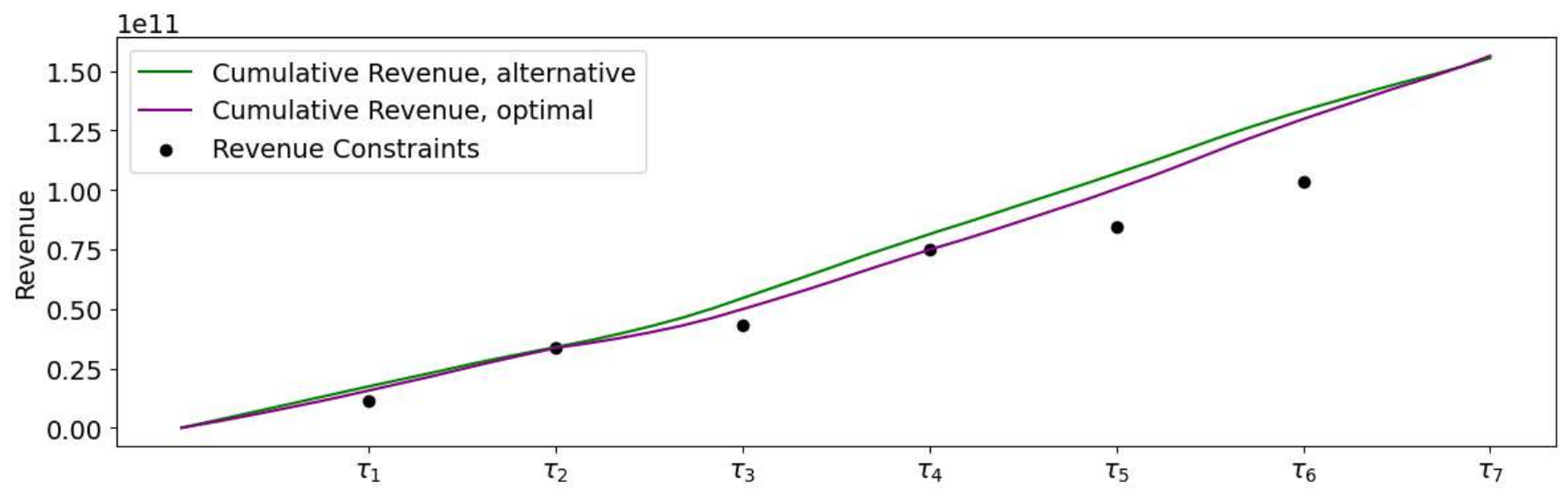}
    \caption{Simulation 2, Cumulative Revenue}
    \label{fig:sim2:revenue}
\end{figure}

\begin{figure}[h!]
    \centering
    \includegraphics[width=1\linewidth]{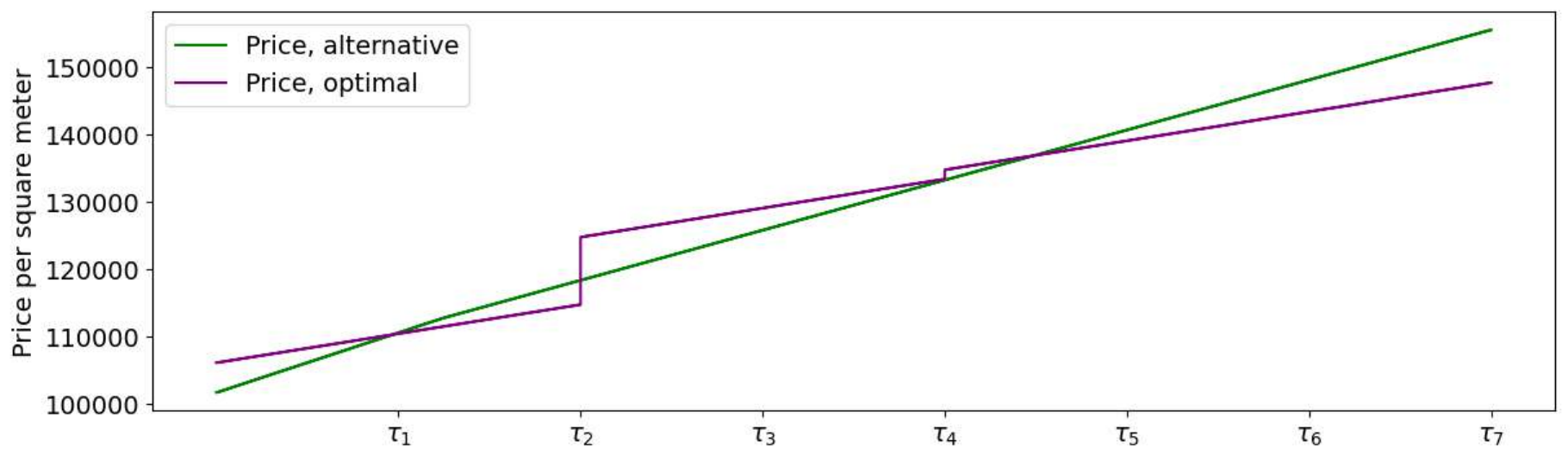}
    \caption{Simulation 2, Price}
    \label{fig:sim2:price}
\end{figure}

The second simulation (see figures~\ref{fig:sim2:sales},\ref{fig:sim2:revenue},\ref{fig:sim2:price}) demonstrates the importance of accounting for the TVM function $\varphi(t)$. In this simulation, $\varphi(t)$ decreases exponentially, while $\kappa(t)=1$ remains constant. The alternative algorithm lacks knowledge about $\varphi(t)$, whereas the optimal algorithm incorporates it. Although the difference in pricing policies is smaller compared to the first simulation, it still results in a significant $0.5\%$ difference in revenue.  


\begin{figure}[h!]
    \centering
    \includegraphics[width=1\linewidth]{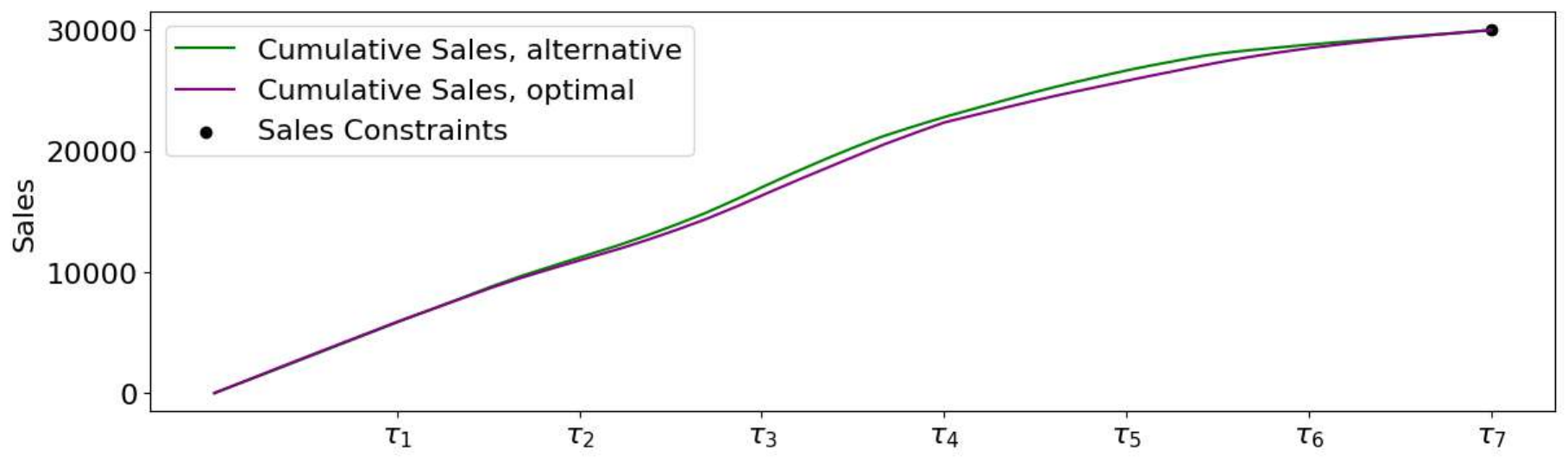}
    \caption{Simulation 3, Cumulative Sales}
    \label{fig:sim3:sales}
\end{figure}

\begin{figure}[h!]
    \centering
    \includegraphics[width=1\linewidth]{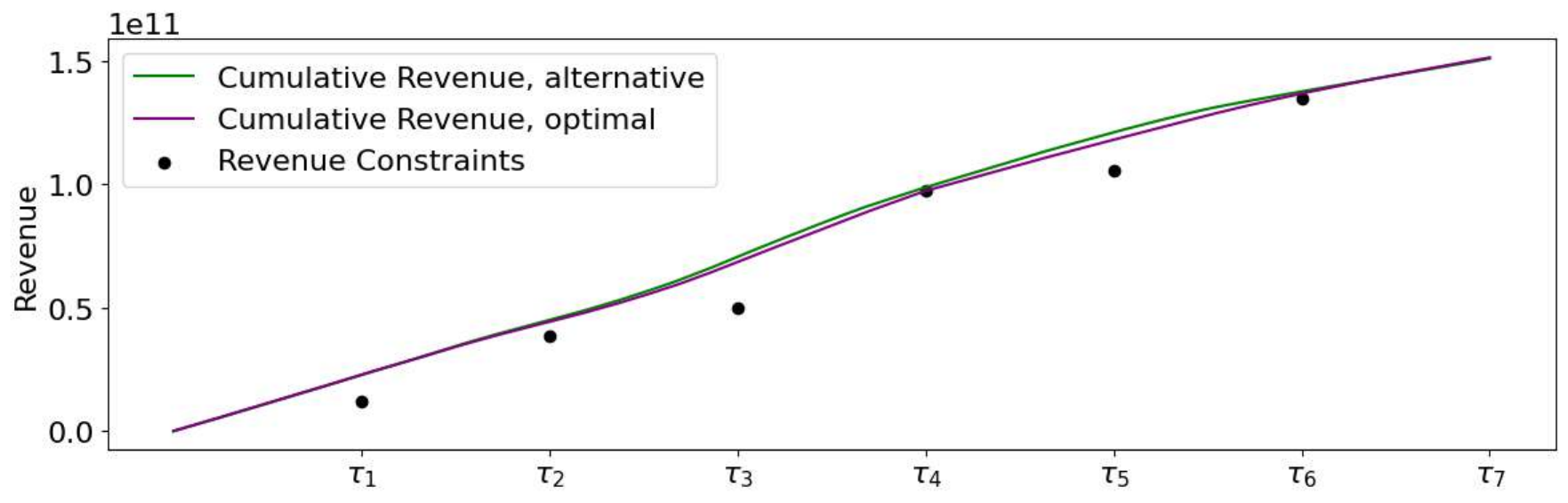}
    \caption{Simulation 3, Cumulative Revenue}
    \label{fig:sim3:revenue}
\end{figure}

\begin{figure}[h!]
    \centering
    \includegraphics[width=1\linewidth]{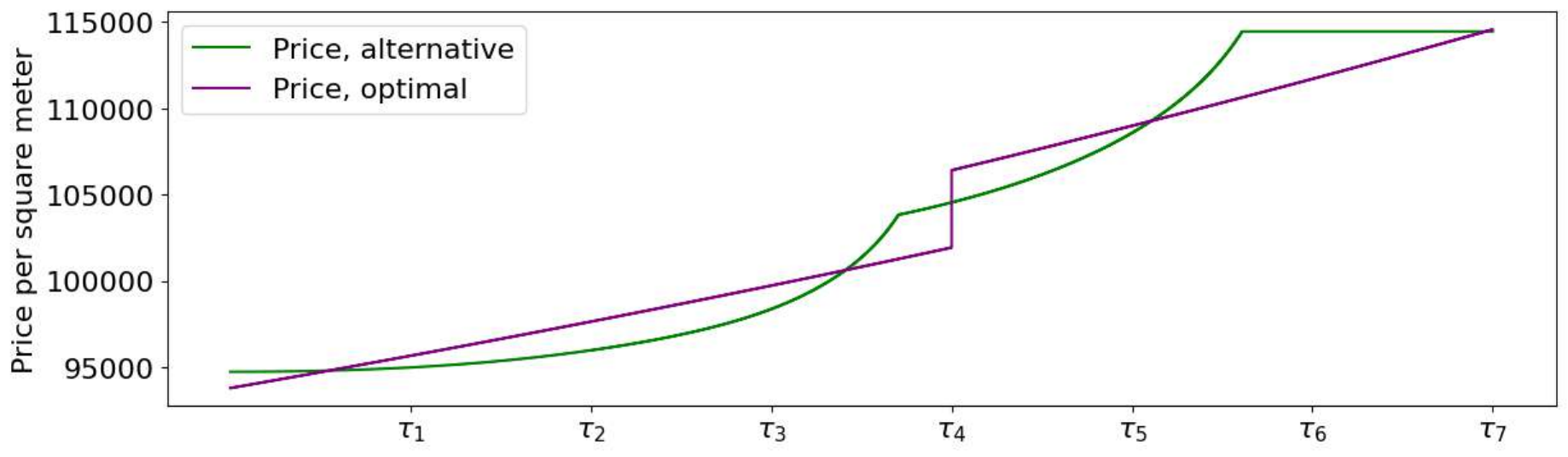}
    \caption{Simulation 3, Price}
    \label{fig:sim3:price}
\end{figure}

Third simulation (see figures~\ref{fig:sim3:sales},\ref{fig:sim3:revenue},\ref{fig:sim3:price}) explores the importance of $\kappa(t)$. We consider a scenario where $\kappa(t)$ increases linearly, while $\varphi(t)=1$ remains constant. The alternative algorithm on step $t$ uses the current $v(p,t)$ but does not account for the increasing trend in purchase propensity. This assumption is quite generous, as in practice, any algorithm would have a reaction delay due to the need to collect sufficient data to update its estimate of purchase propensity, $\widehat{v}(p)$. Similar to the previous simulation, the optimal algorithm yields $0.4\%$ higher revenue. 


Although these simulations test only the optimal theoretical pricing policies, they show that any algorithm that tries to mimic the optimal policy and takes into account $\varphi(t)$ and $\kappa(t)$ will generally yield significantly higher revenue than algorithms based on other principles.


\section{Further directions}

In this article, we considered the basic model and its enhanced versions, which account for variable total demand, the time value of money, and the growth of the objective value of real estate with the development stage, all within a fully deterministic framework -- assuming that both total demand and the purchase propensity functions are known. In this section, we discuss potential avenues for further enhancing this model to make it suitable for practical business applications.

\begin{itemize}
    \item \textit{Multiple typologies}. One of the key assumptions we made is that all real estate objects are homogeneous, which is typically not the case in reality. In practice, there can be multiple pricing groups within the same building, such as one-bedroom, two-bedroom units, and so on. While our models can naturally be generalized to accommodate multiple pricing groups, this extension falls outside the scope of the present paper.
  
    \item \textit{Soft constraints.} 
    At present, we treat all constraints as ``rigid'', meaning the pricing policy must strictly adhere to them. However, in practice, it might be more practical to consider ``soft'' constraints, where a penalty is incurred if certain constraints are not met. This approach introduces changes to the model and optimization problems, which could be worthwhile to explore.
    \item \textit{Total demand prediction.} In this paper we consider total demand as a known function. To fully implement the methods discussed here in real-life scenarios, it is essential to incorporate predictions of future total demand. There is an abundance of factors affecting total demand. For example: economic activity rate, national currency exchange rate, demographics, and mortgage interest rate. To predict the total demand, one can use the full range of machine-learning methods
    and a large array of historical data, not necessarily data from a specific developer, since macroeconomic factors affect all developers similarly.
    \item \textit{Purchase propensity prediction.} In real-world situations, it is also necessary to predict the purchase propensity function $v(p)$ alongside total demand. One possible approach for this is to use a discrete choice model, which is commonly employed to simulate customer behavior. By combining these models with historical data, it is possible to construct real-time predictions of the purchase propensity function $v(p)$. A particularly promising method in this context is to train a discrete choice model using Siamese Neural Networks.
\end{itemize}

\appendix

\bibliography{dynamic_pricing}
\Addresses

\end{document}